\documentclass[a4paper,11pt]{article}

\usepackage{amsmath}
\usepackage{amssymb}
\usepackage{amsfonts}
\usepackage{graphicx}
\usepackage[colorlinks,citecolor=blue,urlcolor=blue,linkcolor=blue]{hyperref}

\begin{document}

\title{Theoretical analysis of direct $CP$ violation and differential decay width in $D^\pm\to \pi^\pm \pi^+\pi^-$ in phase space around the resonances $\rho^0(770)$ and $f_0(500)$}

\author{ Zhen-Hua~Zhang$^{1, 2, a}$, Ren~Song$^{1}$, Yu-Mo~Su$^{1}$, Gang~L\"{u}$^{3, b}$, and Bo~Zheng$^{1,2,c}$
\\ $^{1}$School of Nuclear Science and Technology, University of South China, \\ Hengyang, Hunan 421001, China.
\\ $^{2}$Cooperative Innovation Centre for Nuclear Fuel Cycle Technology \& Equipment,\\ University of South China, Hengyang, Hunan 421001, China.
\\$^{3}$ College of Science, Henan University of Technology, Zhengzhou 450001, China.
\\ email: $^{a}$\texttt {zhangzh@usc.edu.cn}, $^{b}$\texttt{ganglv@haut.edu.cn}, $^{c}$\texttt{zhengb@ihep.ac.cn}}

\maketitle

\begin{abstract}
We perform a theoretical study on direct $CP$ violation in $D^\pm\to \pi^\pm \pi^+\pi^-$ in phase space around the intermediate states $\rho^0(770)$ and $f_0(500)$.
The possible interference between the amplitudes corresponding to the two resonances is taken into account, and the relative strong phase of the two amplitudes is treated as a free parameter.
Our analysis shows that by properly chosen the strong phase, both the $CP$ violation strength and differential decay width accommodate to the experimental results. 
\end{abstract}

%
\section{Introduction \label{Introduction}}
%
Charge-Parity ($CP$) violation has gain extensive attentions ever since its first discovery in $K^0-\overline{K^0}$ systems in 1964 \cite{Christenson:1964fg}.
Within the Standard Model (SM), $CP$ violation is originated from a complex phase in Cabibbo-Kobayashi-Maskawa (CKM) matrix, which describes the mixing of weak and mass eigenstates of quarks \cite{Kobayashi:1973fv}.

Although it is a small effect in general, $CP$ violation can be relatively large in some decay channels of $B$ and $B_s$ mesons \cite{Carter:1980hr,Bigi:1981qs,Carter:1980tk}.
In fact, large $CP$ violation has been confirmed in some two-body decay channel of $B$ and $B_s$ meson, such as $B^\pm\to D_{CP(+1)}\pi^\pm$ \cite{Aaij:2012kz}, $B^\pm\to \rho^0 K^\pm$ \cite{Aubert:2008bj}, $B^\pm\to\rho^0 K^\ast(892)^\pm$ \cite{delAmoSanchez:2010mz}, $B^\pm\to f^0(1370) \pi^\pm$\cite{Aubert:2009av},  $B^0\to \rho^-K^+$ \cite{BABAR:2011ae}, and $B_s\to \pi^+K^-$ \cite{Aaltonen:2011qt,Aaij:2013iua}.
In recent years, even larger $CP$ violation which localized in three-body decay phase space  of $B$ meson was observed by LHCb collaboration in channels such as $B^\pm\to\pi^\pm\pi^+\pi^-$ and $B^\pm\to K^\pm\pi^+\pi^-$ \cite{Aaij:2013sfa,Aaij:2013bla}.
In view of its anisotropy property for small invariant mass of $\pi^+\pi^-$ pair (no larger than the mass of $\rho^0(770)$) in phase space, the large localized $CP$ violation was first interpreted as a consequence of the interference of the decay amplitudes corresponding to nearby resonances with different spins \cite{Zhang:2013oqa,Zhang:2013iga}.
Some other explanations such as the re-scattering effects of final states \cite{Bhattacharya:2013cvn,Bediaga:2013ela} were also proposed thereafter.

Since it is believed to be very small within SM, $CP$ violation in the charm sector provides a good place for searching for New Physics.
However, due mainly to the pollution of non-perturbative effects of strong interactions, an order of 1\% $CP$ asymmetries  in $D$ meson decay are still understandable in SM \cite{Brod:2012ud,Brod:2011re,Bhattacharya:2012ah}.
To date, no $CP$ violation in charm sector is established.
Though there were some hints of $CP$ violation for the channels $D\to\pi\pi$ and $D\to KK$ \cite{Aaij:2013bra,Collaboration:2012qw}, the latest result from LHCb collaboration showed however, no evidence of $CP$ violation in these channels \cite{Aaij:2014gsa}.

A measurement of $CP$ violation in the three-body decay $D^\pm\to \pi^\pm\pi^+\pi^-$ has also been performed by LHCb \cite{Aaij:2013jxa}.
With very high statistics, no localized or overall $CP$ asymmetries are found.
As has been shown in some aforementioned three-body decay channels of $B$ meson, localized $CP$ asymmetries in phase space can be enhanced by the interference of the decay amplitudes corresponding to two intermediate resonances with different spins, whereas the same mechanism should also apply to $D$ meson decays.
In this paper, we will perform a theoretical analysis of $CP$ violation and differential decay width in the channel $D^\pm\to \pi^\pm\pi^+\pi^-$, and will pay special attention to effects caused by the interference of intermediate state $\rho^0(770)$ and $f_0(500)$.

%
\section{\label{Decay amplitudes} Formalism of decay amplitudes and $CP$ asymmetries}
%
In the region of the phase space around the resonances $\rho^0(770)$ and $f_0(500)$, the process $D^\pm\to\pi^\pm\pi^+\pi^-$ is dominated by two cascade decays, $D^\pm\to\pi^\pm f_0(500)\to\pi^\pm\pi^+\pi^-$ and $D^\pm\to\pi^\pm \rho^0(770)\to\pi^\pm\pi^+\pi^-$. 
For the two weak decays $D^\pm\to\pi^\pm f_0(500)$ and $D^\pm\to\pi^\pm \rho^0(770)$, the corresponding effective Hamiltonian can be expressed as  \cite{Buchalla:1995vs,Guo:1999ip}
\begin{equation}
H_{\Delta C=1}=\frac{G_F}{\sqrt{2}}\left\{\left[\sum_{q=d,s}V_{uq}V_{cq}^{\ast}\left(c_1 O^q_1 + c_2 O^q_2\right)\right]-V_{ub}V_{cb}^{\ast}\sum_{i=3}^{6} c_i O_i\right\}+\text{h.c.},
\end{equation}
where $G_{F}$ is the Fermi constant, $V_{q_1q_2}$ ($q_1$ and $q_2$ represent quarks) is the CKM matrix element, $c_{i}$ ($i=1,\cdots,6$)  is the Wilson coefficient, and $O_i$ is the four quark operator, which can be written as
\begin{eqnarray}
O_{1}^{q}& = &\bar{u}_{\alpha} \gamma_{\mu}(1-\gamma{_5})q_{\beta}\bar{q}_{\beta} \gamma^{\mu}(1-\gamma{_5})
c_{\alpha}\ , \nonumber \\
 O_{2}^{q}& = &\bar{u} \gamma_{\mu}(1-\gamma{_5})q\bar{q} \gamma^{\mu}(1-\gamma{_5})c\ ,  \nonumber \\
O_{3}& =& \bar{u} \gamma_{\mu}(1-\gamma{_5})c \sum_{q\prime}\bar{q}^{\prime}\gamma^{\mu}(1-\gamma{_5})
q^{\prime}\ ,\nonumber \\
O_{4}& =&\bar{u}_{\alpha} \gamma_{\mu}(1-\gamma{_5})c_{\beta}
\sum_{q\prime}\bar{q}^{\prime}_{\beta}\gamma^{\mu}(1-\gamma{_5})q^{\prime}_{\alpha}\ , \nonumber \\
O_{5}& =&\bar{u} \gamma_{\mu}(1-\gamma{_5})c \sum_{q'}\bar{q}^
{\prime}\gamma^{\mu}(1+\gamma{_5})q^{\prime}\ , \nonumber \\
O_{6}& =&\bar{u}_{\alpha} \gamma_{\mu}(1-\gamma{_5})c_{\beta}
\sum_{q'}\bar{q}^{\prime}_{\beta}\gamma^{\mu}(1+\gamma{_5})q^{\prime}_{\alpha}\ , 
\end{eqnarray}
with $O_{1}^{q}$ and $O_{2}^{q}$ being tree operators, $O_{3}-O_{6}$ being QCD penguin operators, $\alpha$ and $\beta$ being color indices, and $q'$ running through all the light flavour quarks.

The effective Hamiltonian for the strong decays $\rho^0\to\pi^+\pi^-$ and $f_0(500)\to\pi^+\pi^-$ can be formally expressed as
\begin{eqnarray}
\mathcal{H}_{\rho^0\pi\pi}&=&ig_{\rho\pi\pi}\rho^0_{\mu}(\pi^-\partial^\mu\pi^+-\pi^+\partial^\mu\pi^-),\\
\mathcal{H}_{f_0\pi\pi}&=&g_{f_0\pi\pi}f_0(2\pi^+\pi^-+\pi^0\pi^0),
\end{eqnarray}
where $\rho^0_\mu$, $f_0$ and $\pi^\pm$ are the field operators for $\rho^0$, $f_0(500)$ and $\pi$ mesons, respectively, $g_{\rho\pi\pi}$ and $g_{f_0\pi\pi}$ are the effective coupling constants, which can be expressed in terms of the decay widths as
\begin{eqnarray}
g_{\rho\pi\pi}^2&=&\frac{48\pi}{\Big(1-\frac{4m_\pi^2}{m_\rho^2}\Big)^{3/2}}\cdot\frac{\Gamma_{\rho^0\to\pi^+\pi^-}}{m_\rho},\\
g_{f_0\pi\pi}^2&=&\frac{4\pi m_{f_0}\Gamma_{f_0\to\pi^+\pi^-}}{\Big(1-\frac{4m_\pi^2}{m_{f_0}^2}\Big)^{1/2}}.
\end{eqnarray}
Both $f_0(500)$ and $\rho^0(770)$ decay into one pion pair dominantly through strong interaction, and the isospin symmetry of the strong interaction tells us that $\Gamma_{\rho^0}\simeq\Gamma_{\rho^0\to\pi^+\pi^-}$, and $\Gamma_{f_0}\simeq\frac{3}{2}\Gamma_{f_0\to\pi^+\pi^-}$.

The decay amplitudes for the cascade decays $D^+\to\pi^+ f_0(500)\to\pi^+\pi^+\pi^-$ and $D^+\to\pi^+ \rho^0(770)\to\pi^+\pi^+\pi^-$ take the form
\begin{eqnarray}
\mathcal{M}_{\rho^0}(s_{\text{low}},s_{\text{high}})&=&\frac{\sqrt{2}G_F g_{\rho\pi\pi}m_\rho(s_{\text{high}}-\Sigma_{s_{\text{low}}})}{s_{\text{low}}-m_{\rho^2}+im_\rho\Gamma_\rho}\cdot\Big\{ V_{ud}V_{cd}^\ast\left(-\frac{1}{\sqrt{2}}a_1f_\rho F_1+a_2f_\pi A_0\right)
\nonumber\\&&
-V_{ub}V_{cb}^\ast\Big[\Big(a_4-\frac{2m_\pi^2a_6}{(m_c+m_d)(m_u+m_d)}\Big)f_\pi A_0\Big]\Big\},\\
\mathcal{M}_{f_0}(s_{\text{low}},s_{\text{high}})&=&\frac{\sqrt{2}G_F  (m_D^2-m_\pi^2)g_{f_0\pi\pi}f_\pi F_0}{s_{\text{low}}-m_{f_0^2}+im_{f_0}\Gamma_{f_0}}
\cdot\Big\{ V_{ud}V_{cd}^\ast a_2
\nonumber\\&&
-V_{ub}V_{cb}^\ast\Big[a_4-\frac{2m_\pi^2a_6}{(m_c+m_d)(m_u+m_d)}\Big]\Big\},
\end{eqnarray}
respectively, where $s_{\text{low}}$ and $s_{\text{high}}$ are the invariant mass squared of $\pi^+\pi^-$ pairs with lower and higher invariant masses, respectively, 
$\Sigma_{s_{\text{low}}}=(s_{\text{high,max}}+s_{\text{high,min}})/2$, with $s_{\text{high,max(min)}}$ being the maximum (minimum) value of $s_{\text{high}}$ allowed by phase space for each $s_{\text{low}}$,
 $F_0$, $F_1$ and $A_0$ are short for the form factors $F_0^{D\to f_0}(m_\pi^2)$, $F_1^{(D\to \pi)}(s_{\text{low}})$ and $A_0^{(D\to \rho)}(m_\pi^2)$, respectively.
 All the $a_i$'s are built up from the Wilson coefficients $c_{i}$'s, and take the form $a_i=c_i+c_{i+1}/N_c$ for odd $i$ and $a_i=c_i+c_{i-1}/N_c$ for even $i$.

In the phase space that we are considering, the total decay amplitude for $D^+\to \pi^+\pi^+\pi^-$ is dominated by $\mathcal{M}_{f_0}$ and $\mathcal{M}_{\rho^0}$.
As a result, it can be expressed as
\begin{equation}\label{TotalAmplitude}
\mathcal{M}=\left[\mathcal{M}_{f_0}(s_{\text{low}},s_{\text{high}})+\mathcal{M}_{\rho^0}(s_{\text{low}},s_{\text{high}})e^{i\delta}\right]+[s_{\text{low}}\leftrightarrow s_{\text{high}}],
\end{equation}
where $\delta$ is the relative strong phase between the two amplitudes $\mathcal{M}_{f_0}$ and $\mathcal{M}_{\rho^0}$, which in principle, arises from long distance effect, $[s_{\text{low}}\leftrightarrow s_{\text{high}}]$ represents a term which take the same form as that in the first square bracket except an interchange between $s_{\text{low}}$ and $s_{\text{high}}$.
For the calculation of the decay amplitude of the $CP$ conjugate process $D^-\to\pi^+\pi^-\pi^-$, which will be denoted as $\overline{\mathcal{M}}$, all one need to do is to replace the CKM matrix elements in $\mathcal{M}$ with their complex conjugates.

The differential $CP$ asymmetry for $D^\pm\to\pi^\pm\pi^+\pi^-$ is defined as
\begin{equation}
A_{CP}=\frac{|\mathcal{M}|^2-|\overline{\mathcal{M}}|^2}{|\mathcal{M}|^2+|\overline{\mathcal{M}}|^2},
\end{equation}
while the localized $CP$ asymmetry can be expressed as
\begin{equation}
A_{CP}^{R}=\frac{\int_R ds_{\text{high}}ds_{\text{low}}(|\mathcal{M}|^2-|\overline{\mathcal{M}}|^2)}{\int_R ds_{\text{high}}ds_{\text{low}}(|\mathcal{M}|^2+|\overline{\mathcal{M}}|^2)},
\end{equation}
where, $R$ represents certain region of the phase space that we are considering.

%
\section{\label{Numerical analysis}Numerical analysis}
%
Table 1 list the input parameters and the corresponding references we used in this paper.
In the following, we give some comments on these input parameters.
We use the Wolfenstein parameterization for the CKM matrix elements, which up to the order of $\lambda^8$ , can be expressed as \cite{Buras:1994ec,Charles:2004jd}
\begin{eqnarray}
V_{ud}&=&1-\frac{\lambda^2}{2}-\frac{\lambda^4}{8}-\frac{\lambda^6}{16}[1+8A^2(\rho^2+\eta^2)]-\frac{\lambda^8}{128}[5-32A^2(\rho^2+\eta^2)],\nonumber\\
V_{cd}&=&-\lambda+\frac{\lambda^5}{2}A^2[1-2(\rho+i\eta)]+\frac{\lambda^7}{2}A^2(\rho+i\eta),\nonumber\\
V_{ub}&=&\lambda^3A(\rho-i\eta),\nonumber\\
V_{cb}&=&A\lambda^2-\frac{\lambda^8}{2}A^3(\rho^2+\eta^2),
\end{eqnarray}
with $A$, $\rho$, $\eta$, and $\lambda$ being the Wolfenstein parameters.
To all orders in $\lambda$, the relation between $\rho$, $\eta$ and $\overline{\rho}$, $\overline{\eta}$ can be expressed as \cite{Charles:2004jd}
\begin{equation}
\rho+i\eta=\frac{\sqrt{1-A^2\lambda^4}(\overline{\rho}+i\overline{\eta})}{\sqrt{1-\lambda^2}[1-A^2\lambda^4(\overline{\rho}+i\overline{\eta})]}.
\end{equation}
For the invariant mass dependence of the form factors $F_1^{D\to \pi}$ and $A_0^{D\to\rho}$, we use a model from Ref. \cite{Cheng:2003sm}, which take the form 
\begin{equation}
F(s)=\frac{F(0)}{1-a^X\cdot\frac{s}{m_D^2}+b^X\cdot(\frac{s}{m_D^2})^2},
\end{equation}
where $F$ and $X$ can be $F_1^{D\to \pi}$ and  $\pi$, or $A_0^{D\to\rho}$ and $\rho$, respectively.
The form factor $F^{D\to f_0}(m_{\pi}^2)$  which we use here is a rough estimation, and is consistent with branching ratio of $D^+\to f_0(500)\pi^+$ extracted from Dalitz analysis of Data \cite{Bonvicini:2007tc}.

\begin{table}[tbp]
\caption{\label{InputParameters}Input parameters used in this paper.}
\begin{tabular}{|*{8}{l|}}
\hline
Parameters&Input data & References\\
\hline
Fermi constant (in $\text{GeV}^{-2}$)&$G_F=1.16638\times10^{-5}$& \cite{Agashe:2014kda}\\
\hline
Wilson coefficients&$c_1=-0.6941,~c_2=1.3777,$& \cite{Guo:1999ip}\\
&$c_3=0.0652,~c_4=-0.0627,$&\\
&$c_5=0.0206,~c_6=-0.1355,$&\\
\hline
Masses and decay widths&$m_{D^\pm}=1.86961,~\tau_{D^\pm}=1.040\times10^{-12}s$&\cite{Agashe:2014kda}\\
(in GeV)&$BR(D^+\to\pi^+\pi^-\pi^+)=3.18\times10^{-3}$&\\
&$m_{\rho^0(770)}=0.775, ~\Gamma_{\rho^0(770)}=0.150,$&\\
&$m_{f_0(500)}=0.5, ~\Gamma_{f_0(500)}=0.5,$&\\
&$m_\pi=0.13957,$&\\
&$m_u=0.0023,~m_d=0.0048,$&\\
&$m_s=0.095,~m_c=1.275,$&\\
\hline
Form factors&$F_1^{D\to \pi}(0)=0.67,~A_0^{D\to\rho}(0)=0.64,$&\cite{Cheng:2003sm}\\
&$a^\pi=1.19,~b^\pi=0.36$&\\
&$a^\rho=1.07,~b^\rho=0.54$&\\
\cline{2-3}
&$F_0^{F\to f_0}(m_{\pi}^2)=0.33,$&--\\
\hline
Decay constants&$f_\pi=0.13041,~f_K=0.1562,$&\cite{Agashe:2014kda}\\
\cline{2-3}
(in GeV)&$f_\rho=0.216,$& \cite{Maris:1999nt}\\
\hline
Wolfenstein parameters &$\lambda=0.22548_{-0.00034}^{+0.00068},~A=0.810_{-0.024}^{+0.018},$&\cite{Charles:2015gya}\\
of CKM matrix&$\overline{\rho}=0.145_{-0.007}^{+0.013}, ~\overline{\eta}=0.343_{-0.012}^{+0.011},$&\\
\hline
\end{tabular}
\end{table}

\begin{figure}
\includegraphics[width=\textwidth]{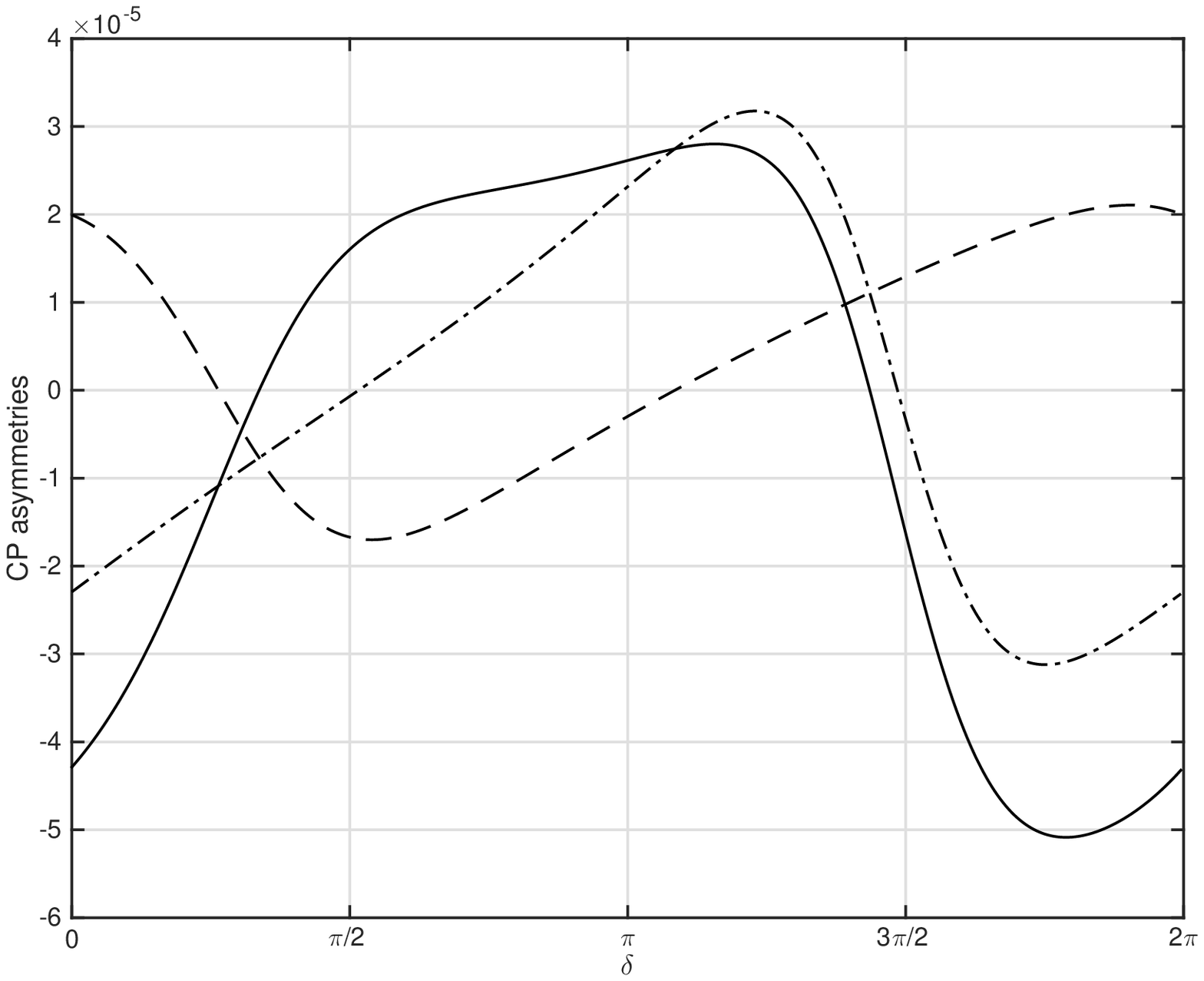}
\caption{\label{DeltaACP} The $CP$ asymmetry difference $\Delta A_{CP}$ (solid line), the $CP$ asymmetry of region $\Omega^-$ (dashed line) and $\Omega^+$ (dash-dotted line) as a function of the strong phase $\delta$.}
\end{figure}

As is observed by LHCb, the $CP$ asymmetries around the vicinities of $f_0(500)$ and $\rho^0(770)$ have opposite signs for small and large values of $s_{\text{high}}$ in the case of $B$ meson decay channel $B^\pm\to\pi^\pm\pi^+\pi^-$ \cite{Aaij:2013bla}. 
In view of the above, for the case of $D^\pm\to\pi^\pm\pi^+\pi^-$, we will focus on $CP$ asymmetries of two regions, denoted as $\Omega^+$ and $\Omega^-$, where $\Omega^+$ ($\Omega^-$) represents  phase space satisfying  $s_{\text{high}}>(<)\Sigma_{s_\text{low}}$, and $s_{\text{high}}>m_\rho^2$.
The $CP$ asymmetry difference of the aforementioned two regions is
\begin{equation}
\Delta A_{CP}=A_{CP}^{\Omega^+}-A_{CP}^{\Omega^-}.
\end{equation}
In Fig.~\ref{DeltaACP}, the $CP$ asymmetries $A_{CP}^{\Omega^+}$, $A_{CP}^{\Omega^-}$ and their difference $\Delta A_{CP}$ are shown as a function of the strong phase $\delta$, where the strong phase $\delta$ is assumed as a constant with respect to $s_{\text{low}}$ and $s_{\text{high}}$.
It can be seen from Fig. \ref{DeltaACP} that $\Delta A_{CP}$ is negative(positive) when $\delta$ is around 0 ($\pi$).
The magnitude of $\Delta A_{CP}$ can reach as large as $0.5\times10^{-4}$ for some values of $\delta$.
Especially, Fig.~\ref{DeltaACP} shows that our mechanism indicates possibilities for  $\Delta A_{CP}$ being zero.
This is interesting because that the experimental result from LHCb collaboration shows no $CP$ violation in this channel.
One can read off two zero points for $\Delta A_{CP}$ from Fig.~\ref{DeltaACP}, which are $\delta_1=4.50$ and $\delta_2=1.06$.

\begin{figure}
\includegraphics[width=\textwidth]{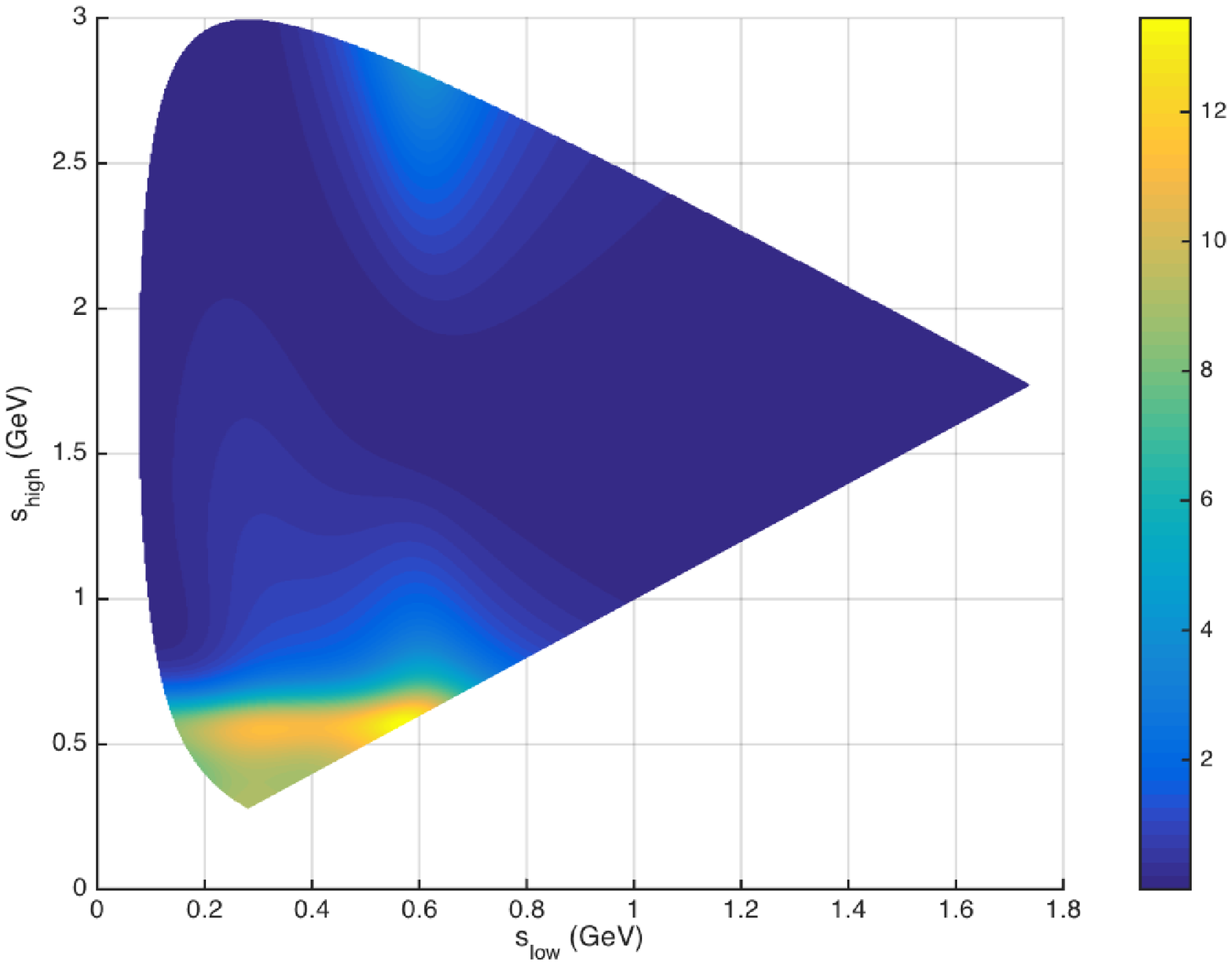}
\caption{\label{DifWidthDelta1} Differential branching ratio (in $\text{GeV}^{-4}$) of $D^+\to \pi^+\pi^-\pi^+$  when $\delta=4.50$.}
\end{figure}

\begin{figure}
\includegraphics[width=\textwidth]{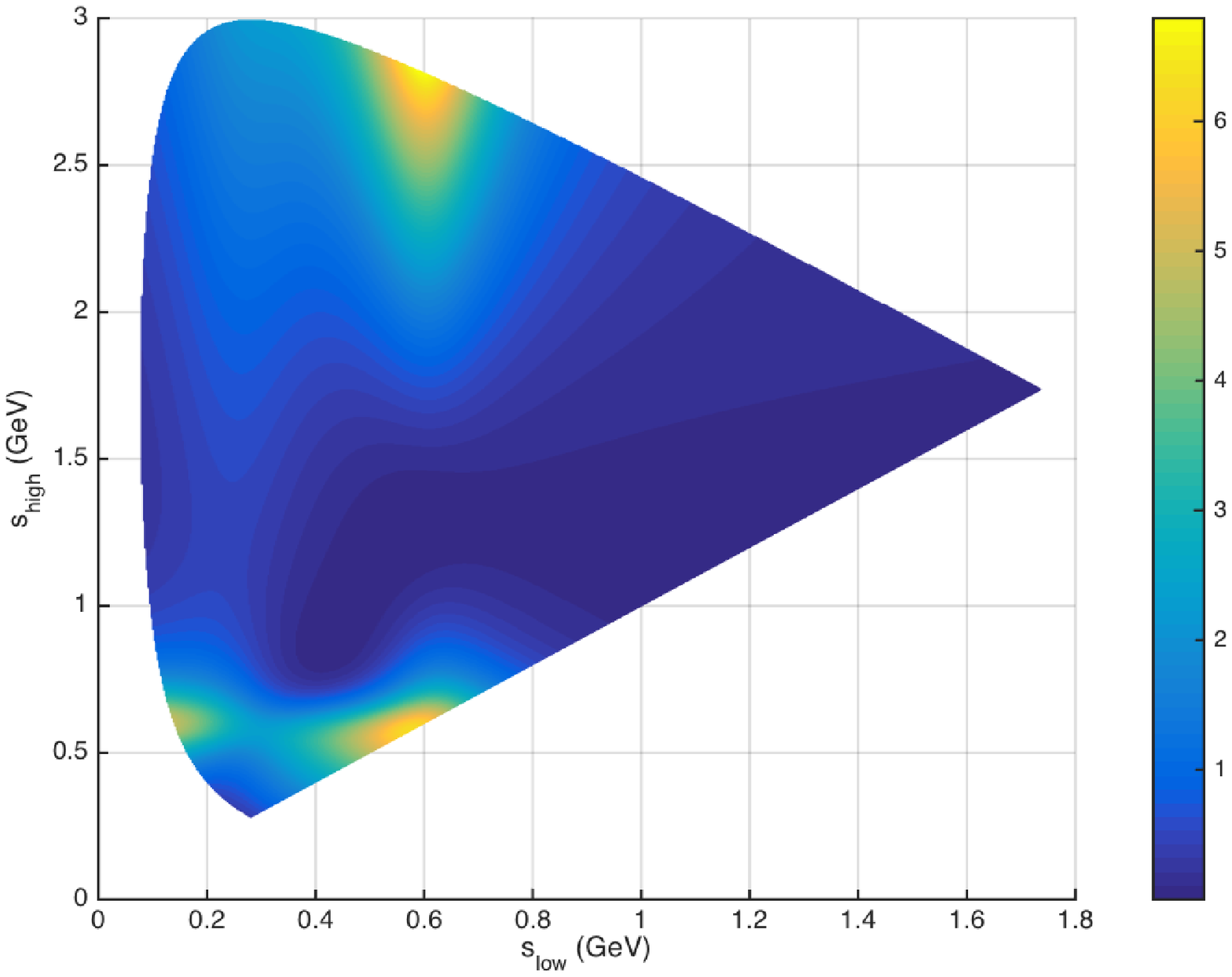}
\caption{\label{DifWidthDelta2} Differential branching ratio (in $\text{GeV}^{-4}$) of $D^+\to \pi^+\pi^-\pi^+$  when $\delta=1.06$.}
\end{figure}

\begin{figure}
\includegraphics[width=\textwidth]{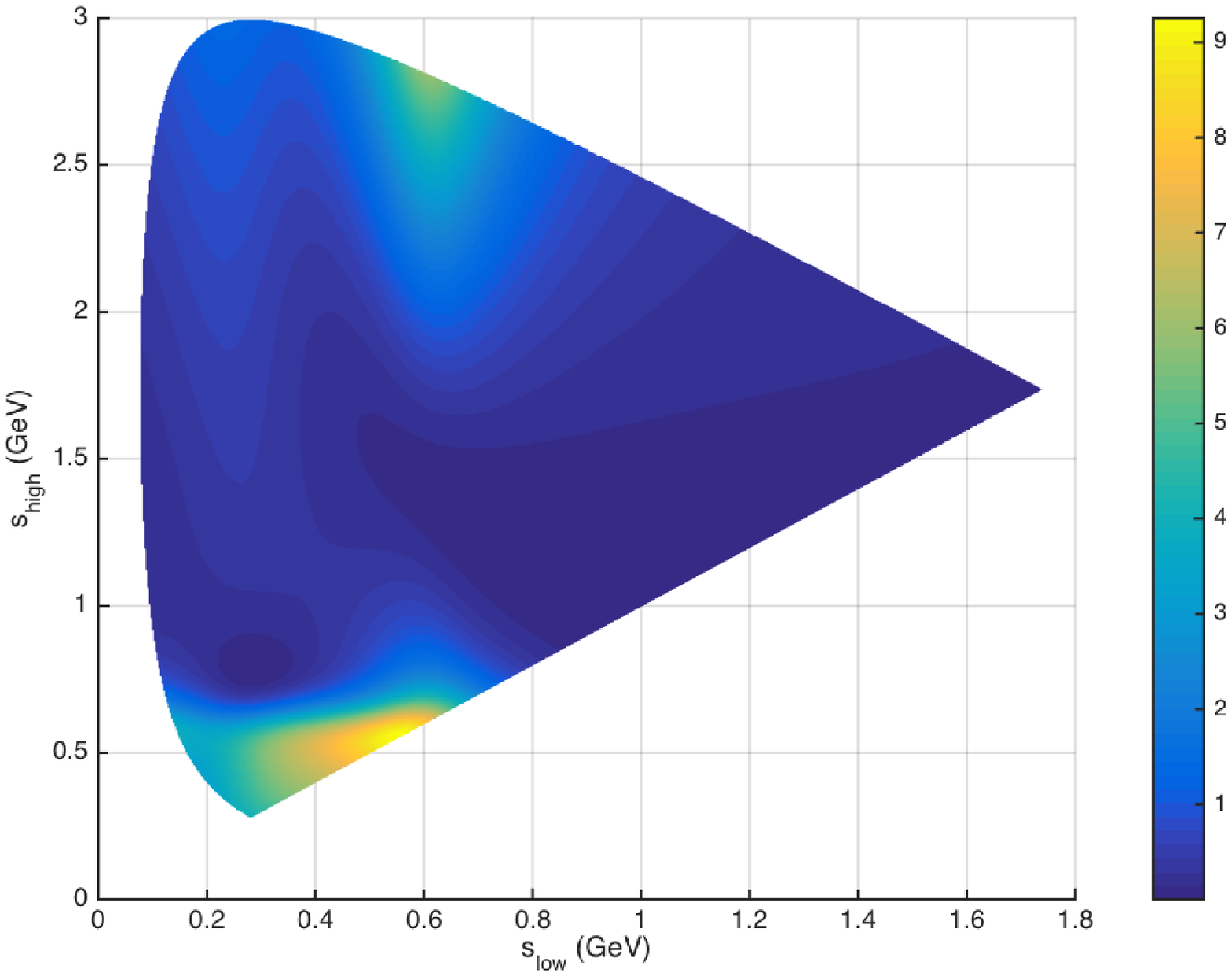}
\caption{\label{DifWidthDeltaPi} Differential branching ratio (in $\text{GeV}^{-4}$) of $D^+\to \pi^+\pi^-\pi^+$  when $\delta=0$.}
\end{figure}

\begin{figure}
\includegraphics[width=\textwidth]{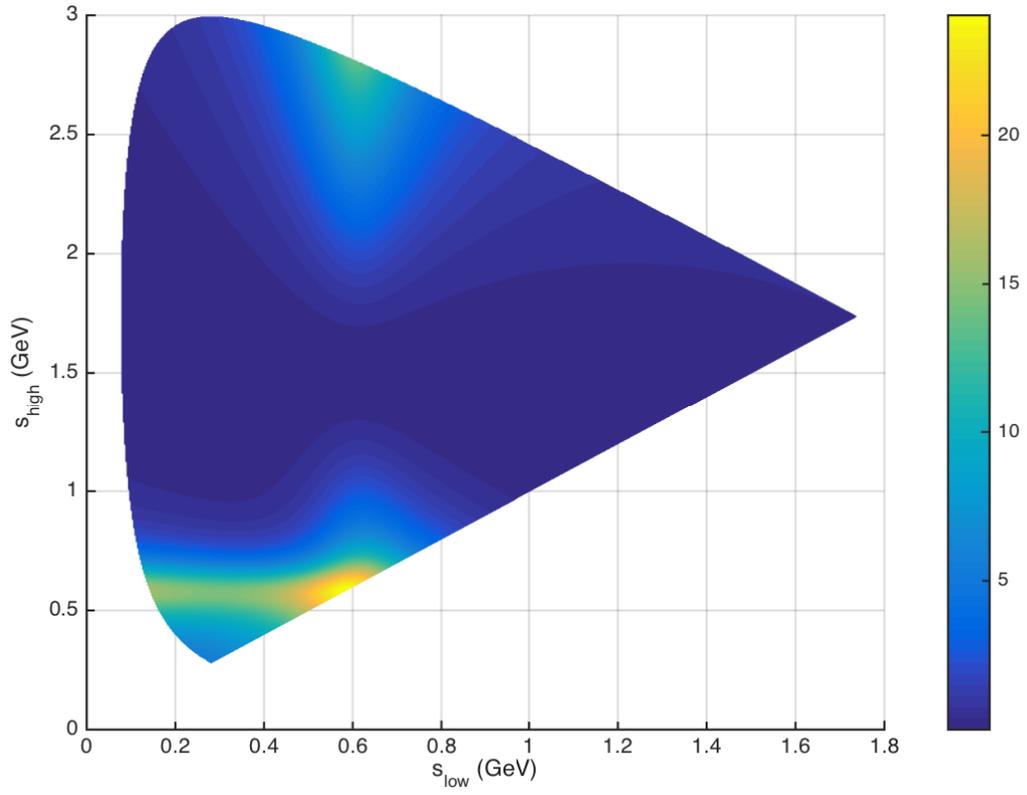}
\caption{\label{DifWidthWithoutInt} Differential branching ratio (in $\text{GeV}^{-4}$) of $D^+\to \pi^+\pi^-\pi^+$,  where only the amplitudes corresponding to $\rho^0(770)$ are included.}
\end{figure}

Figures \ref{DifWidthDelta1} and \ref{DifWidthDelta2} present  in the phase space for $\delta=4.50$ and $\delta=1.06$, respectively, the relative differential decay width $\gamma$ of $D^+\to\pi^+\pi^-\pi^+$,  which is defined as
\begin{equation}
\gamma(s_{\text{low}},s_{\text{high}})\equiv\frac{1}{\Gamma_{D^+\to\pi^+\pi^-\pi^+}}\cdot\frac{d^2\Gamma_{D^+\to\pi^+\pi^-\pi^+}}{ds_{\text{low}}ds_{\text{high}}}=\frac{1}{256\pi^3m_D^3\Gamma_{D^+\to\pi^+\pi^-\pi^+}}\left|\mathcal{M}\right|^2.
\end{equation}
For comparison, we also present the relative differential decay width in Fig.~\ref{DifWidthDeltaPi}  for $\delta=0$, and that in Fig.~\ref{DifWidthWithoutInt} for the situation only resonance $\rho^0(770)$ is taken into account.
Experimental data from LHCb shows that symmetries of event distribution around the $\rho^0(770)$ resonance are badly destroyed.
The number of events around the $\rho^0(770)$ resonance for $s_{\text{high}}<\Sigma_{s_{\text{low}}}$ are much larger than that for $s_{\text{high}}>\Sigma_{s_{\text{low}}}$, as is shown in Ref.~\cite{Aaij:2013jxa}.
Besides, LHCb results also shows an enhancement of event distributions in the region of phase space where $\sqrt{s_{\text{low}}}$ and $\sqrt{s_{\text{high}}}$ are around  the masses of $f_0(500)$ and $\rho^0(770)$, respectively.
These behaviours are roughly the same as those shown in Fig.~\ref{DifWidthDelta1}, which indicate that our mechanism is consistent with experimental data when $\delta=4.50$.

%
\section{\label{Discussion} Discussion}
%
We used a naive factorization approach for the weak decay processes in the calculation of the decay amplitudes.
The reason is simply because that large part of the region in phase space that we focused on is off the mass shells of $\rho^0(770)$ and $f_0(500)$, the advantages of factorization approaches such as perturbative QCD \cite{Li:2001vm}, QCD factorization \cite{Beneke:1999br}, Soft Collinear Effective Theory \cite{Bauer:2001cu} are smeared out by the off shell effect.

In determining the strong phase $\delta$, we used the $CP$ asymmetry difference of two regions of phase space instead of the differential $CP$ asymmetry.
The reason is because that the use of the differential $CP$ asymmetry as a tool to determine the strong phase $\delta$ is not an appropriate approach at all.
On one hand, if use the differential $CP$ asymmetry, one would find that no strong phase can accommodate that with the data.
On the other hand, if one check the nonzero differential $CP$ asymmetries distributed in phase space for $\delta=4.50$, one would find that the large differential $CP$ asymmetries goes always with very small differential decay amplitude $\mathcal{M}$, indicating a cancellation between $\mathcal{M}_{\rho}e^{i\delta}$ and $\mathcal{M}_{f_0}$.
In this situation, the dominance of these two amplitudes is no longer valid. 
Consequently, in order to deduce  the differential $CP$ asymmetries in this kind of regions, we should in principle consider other contributions to the decay amplitude $\mathcal{M}$ in these phase space regions, which is out of the scope of this paper.

We choose the right boundary of the two regions $\Omega^+$ and $\Omega^-$ in phase space to be $m_\rho^2$. 
Although it is not an unique choice, the boundary should not be far away from the vicinity of $\rho^0(770)$, in which case, the allowed strong phase $\delta$ is not sensitive to the choice.
On the other hand, either it is too large or too small than $m_\rho^2$, the dominance of two resonance $\rho^0$ and $f_0(500)$ of the total decay amplitude is no longer valid.

Other resonances can also contribute to the decay amplitude.
For resonance such as $f_0(980)$, only a small part of the total resonance lies in the region. 
As a result, this contribution is small compared with $\rho^0(770)$.
Resonance $\omega$ can enter in the amplitude through an isospin breaking effect, which is called $\rho^0-\omega$ mixing mechanism.
This mechanism can generate large differential $CP$ asymmetries in the vicinity of $\Omega$.
However, the width of $\omega$ is small, its contribution to regional $CP$ violation is small.
More importantly, the contribution of $\omega$ to $CP$ asymmetry is independent of $s_{\text{high}}$, and hence has no contribution to $\Delta A_{CP}$ \footnote{Strictly speaking, since there are two identical particles in the final state for $D^\pm\to\pi^\pm\pi^+\pi^-$, the terms of amplitudes are doubled by $[s_{\text{low}}\leftrightarrow s_{\text{high}}]$, as is shown in Eq. \ref{TotalAmplitude}. As a result, the contribution of resonance $\omega$  in $\Delta A_{CP}$ cannot be cancelled exactly. Besides, the interference between the amplitudes corresponding to $f_0(500)$ and $\omega$ is also small due to the smallness of $\omega$'s width.}.

From Fig.~\ref{DeltaACP}, one can see that the $CP$ asymmetries of the two regions $\Omega^+$ and $\Omega^-$ are nonzero for $\delta=4.50$, which seems to be a disadvantage of our work.
However, these two $CP$ asymmetries are small, which both are $1.1\times10^{-5}$.
In principle, these small $CP$ asymmetries are understandable by, for example, an inclusion of $s_{\text{low}}$ or $s_{\text{high}}$ dependence on $\delta$.
What important is, our division of the phase space $f_0(500)$ and $\rho^0(770)$ enlarge the effect of  $CP$ violations caused by the  interference,
consequently, can be used to determine the strong phase.

%
\section{\label{Conclusion}Conclusion}
%
In this paper, we study the localized $CP$ violation and differential decay width of the decay channel $D^\pm\to\pi^+\pi^-\pi^+$.
We focus our attention on phase space where the invariant mass of $\pi^+\pi^-$ are around the vicinities of $\rho^0(770)$ and $f_0(500)$.
We consider a mechanism which can generate large $CP$ asymmetries on three-body decays of $B$ meson, that is localized $CP$ asymmetries caused by the interference of amplitudes corresponding to resonances with different spins.
We found that by properly choosing a relative strong phase $\delta$, the interference of amplitude corresponding to resonances $f_0(500)$ and $\rho^0(770)$ gives predictions that are consistent with experimental data both on $CP$ asymmetries and differential decay widths. 
Our results generate no $CP$ asymmetry differences ($\Delta A_{CP}=0$) when the strong phase $\delta=4.50$.
In the same time, the behaviour of event distribution around the vicinity of $\rho^0(770)$ and $f_0(500)$ is also understandable.

%
\section*{Acknowledgments}
%
This work was partially supported by National Natural Science Foundation of China (No. 11447021), the construct program of the key discipline in Hunan province, and the Innovation Team of Nuclear and Particle Physics of USC.

\bibliographystyle{unsrt}
 
\bibliography{zzh}


\end{document}